\documentclass[sigconf]{acmart}
\usepackage{algpseudocode}
\usepackage{listings}
\usepackage{bm}
\usepackage[ruled,vlined,linesnumbered]{algorithm2e}
\usepackage{subfigure}
\usepackage{xcolor}
\usepackage{graphicx}
\usepackage{multirow}
\usepackage{amsmath,amsfonts}

\AtBeginDocument{%
  }

\copyrightyear{2025}
\acmYear{2025}
\setcopyright{acmlicensed}\acmConference[MM '25]{Proceedings of the 33rd ACM International Conference on Multimedia}{October 27--31, 2025}{Dublin, Ireland}
\acmBooktitle{Proceedings of the 33rd ACM International Conference on Multimedia (MM '25), October 27--31, 2025, Dublin, Ireland}
\acmDOI{10.1145/3746027.3763784}
\acmISBN{979-8-4007-2035-2/2025/10}

\begin{document}

\title{Cross-Modal Prototype Augmentation and Dual-Grained Prompt Learning for Social Media Popularity Prediction}


\author{Ao Zhou}
\authornote{Equal contribution.}
\affiliation{%
  \institution{State Key Laboratory for Novel Software Technology, Nanjing University}
  \city{Nanjing}
  \country{China}
}\email{zacqupt@gmail.com}

\author{Mingsheng Tu}
\authornotemark[1]
\affiliation{%
  \institution{Chongqing University of Posts and Telecommunications}
  \city{Chongqing}
  \country{China}}
\email{tumingsheng25@gmail.com}

\author{Luping Wang}
\affiliation{%
  \institution{Chongqing University of Posts and Telecommunications}
  \city{Chongqing}
  \country{China}}
\email{s240233001@stu.cqupt.edu.cn}

\author{Tenghao Sun}
\affiliation{%
  \institution{Chongqing University of Posts and Telecommunications}
  \city{Chongqing}
  \country{China}}
\email{kudohao@163.com}

\author{Zifeng Cheng}
\affiliation{%
  \institution{State Key Laboratory for Novel Software Technology, Nanjing University
}
  \city{Nanjing}
  \country{China}
}
\email{chengzf@smail.nju.edu.cn}

\author{Yafeng Yin}
\affiliation{%
  \institution{State Key Laboratory for Novel Software Technology, Nanjing University}
  \city{Nanjing}
  \country{China}
}
\email{yafeng@nju.edu.cn}

\author{Zhiwei Jiang}
\authornote{Corresponding author.}
\affiliation{%
  \institution{State Key Laboratory for Novel Software Technology, Nanjing University
}
  \city{Nanjing}
  \country{China}}
\email{jzw@nju.edu.cn}

\author{Qing Gu}
\affiliation{%
  \institution{State Key Laboratory for Novel Software Technology, Nanjing University}
  \city{Nanjing}
  \country{China}}
\email{guq@nju.edu.cn}

\renewcommand{\shortauthors}{Ao Zhou et al.}

\begin{abstract}
Social Media Popularity Prediction is a complex multimodal task that requires effective integration of images, text, and structured information. 
However, current approaches suffer from inadequate visual-textual alignment and fail to capture the inherent cross-content correlations and hierarchical patterns in social media data. 
To overcome these limitations, we establish a multi-class framework , introducing hierarchical prototypes for structural enhancement and contrastive learning for improved vision-text alignment. Furthermore, we propose a feature-enhanced framework integrating dual-grained prompt learning and cross-modal attention mechanisms, achieving precise multimodal representation through fine-grained category modeling. 
Experimental results demonstrate state-of-the-art performance on benchmark metrics, establishing new reference standards for multimodal social media analysis.
\end{abstract}





\maketitle

\section{Introduction}
The rapid development of mobile internet has deeply integrated social media platforms into all aspects of modern life, fundamentally reshaping human communication and social interaction patterns. With the exponential growth of multimodal user-generated content on platforms like TikTok and Instagram, social posts containing rich media formats such as images, text, and micro-videos are being created and disseminated at a rate of millions per second \cite{SMPapply1,SMPapply2,SMPapply3}. 
This content explosion has made accurate Social Media Popularity Prediction (SMPP) \cite{SMP1,SMP2,SMP3} a research topic of significant theoretical value and practical importance—it not only relates to personal digital influence building and corporate marketing strategy formulation, but also provides decision-making support for critical applications like public opinion monitoring and recommendation system optimization. The fundamental challenges in Social Media Popularity Prediction research originate from the inherent multimodal complexity of social media content, which differs significantly from traditional unimodal analysis \cite{unimodal}. Typical social media posts concurrently integrate multiple information dimensions, including visual elements, textual descriptions, and spatiotemporal metadata. These heterogeneous modalities demonstrate potential semantic correlations while frequently exhibiting expression biases. 
% \begin{figure}[htbp]
% \centering
% \begin{minipage}[t]{0.45\linewidth}
% \centering
% \begin{tabular}{|c|p{0.45\linewidth}|}
% \hline
% \multicolumn{2}{|c|}{\textbf{Post (a)}} \\
% \hline
% \textbf{Photo} & \includegraphics[width=0.98\linewidth]{post1.png} \\
% \hline
% \textbf{Category} & Weather \& Season \\
% \hline
% \textbf{Subcategory} & Raining \\
% \hline
% \textbf{Title} & \textcolor{red}{Puddle} \\
% \hline
% \textbf{All tags} & California longexposure light reflection night puddle nikon parking lot eureka d800e \\
% \hline
% \textbf{Popularity} & 8.46 \\
% \hline
% \end{tabular}
% \end{minipage}
% \hfill
% \begin{minipage}[t]{0.45\linewidth}
% \centering
% \begin{tabular}{|c|p{0.45\linewidth}|}
% \hline
% \multicolumn{2}{|c|}{\textbf{Post (b)}} \\
% \hline
% \textbf{Photo} & \includegraphics[width=0.98\linewidth]{post2.png} \\
% \hline
% \textbf{Category} & Weather \& Season \\
% \hline
% \textbf{Subcategory} & Summer \\
% \hline
% \textbf{Title} & \textcolor{red}{NULL} \\

% \hline
% \textbf{All tags} & Road city travel trees light summer sky mountain lake mountains tree green nature water\\
% \hline
% \textbf{Popularity} & 4.32 \\
% \hline
% \end{tabular}
% \end{minipage}
% \caption{Samples of social media posts.}
% \label{fig:example}
% \end{figure}

Current research on social media popularity prediction predominantly follows a two-stage pipeline consisting of feature extraction followed by regression modeling. The feature extraction approaches generally fall into two categories: (1) traditional feature engineering \cite{Featureselect1,Featureselect2,Featureselect3,Featureselect4} that manually designs various features including user profile attributes, image descriptors, and time-series patterns, which while interpretable, requires substantial domain expertise and suffers from limited generalization capability; and (2) Modern representation learning leverages Language Models (LM) \cite{Bert,T5} and Vision-Language Models (VLM) \cite{CLIP,BLIP} approaches for feature discovery. Notable examples include the Dual-Stream Pre-trained Transformer (DSPT) \cite{DSPT} framework that processes visual and textual data through separate VLM and LM branches, and High-level Vision-Language Alignment (HVLA) \cite{HVLA} approaches utilizing BLIP-2 \cite{BLIP2} for text-guided image captioning to achieve unified semantic space mapping.

\begin{figure}[!h]
    \centering
    \subfigure[Title]{\includegraphics[width=0.23\textwidth]{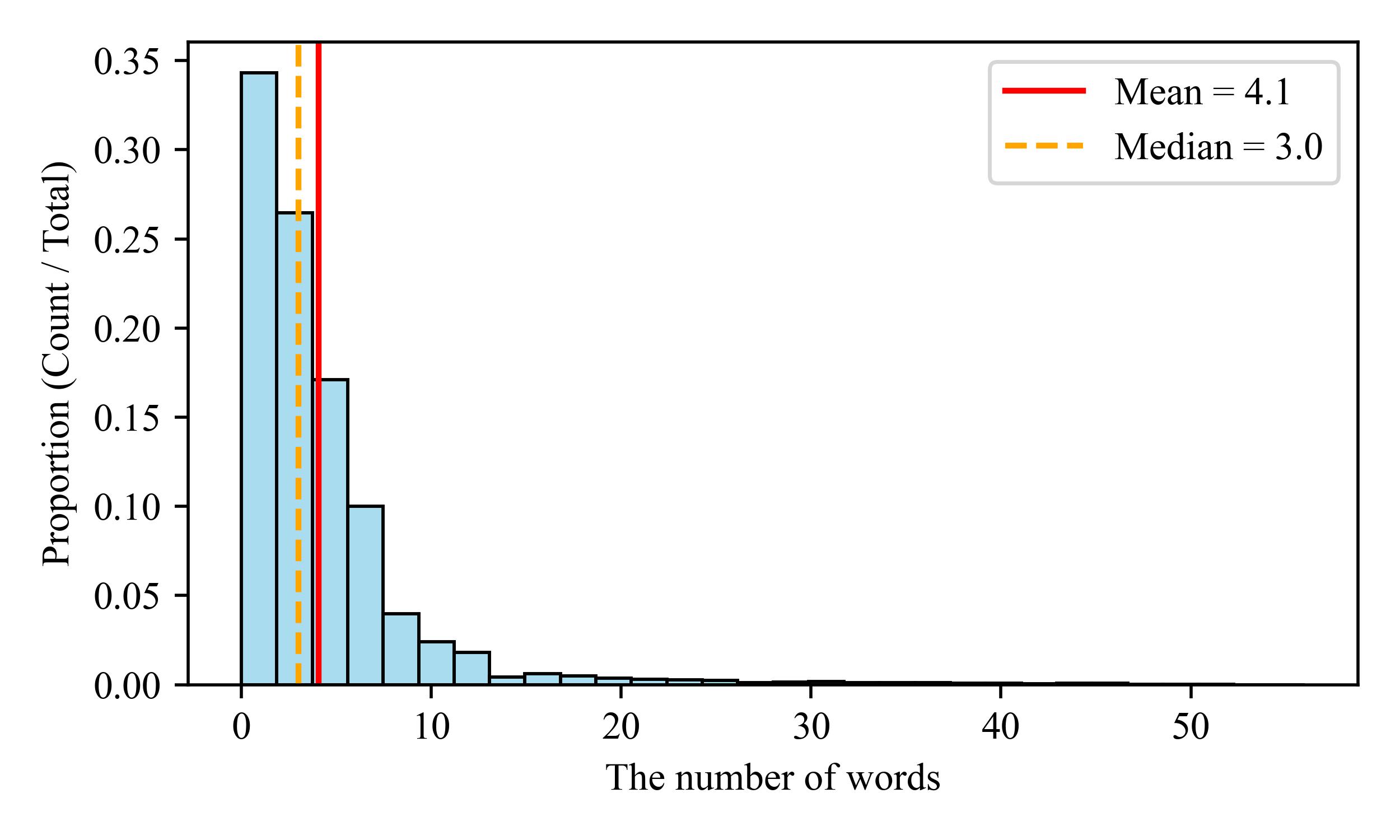}
    }
    \subfigure[All tags]{\includegraphics[width=0.23\textwidth]{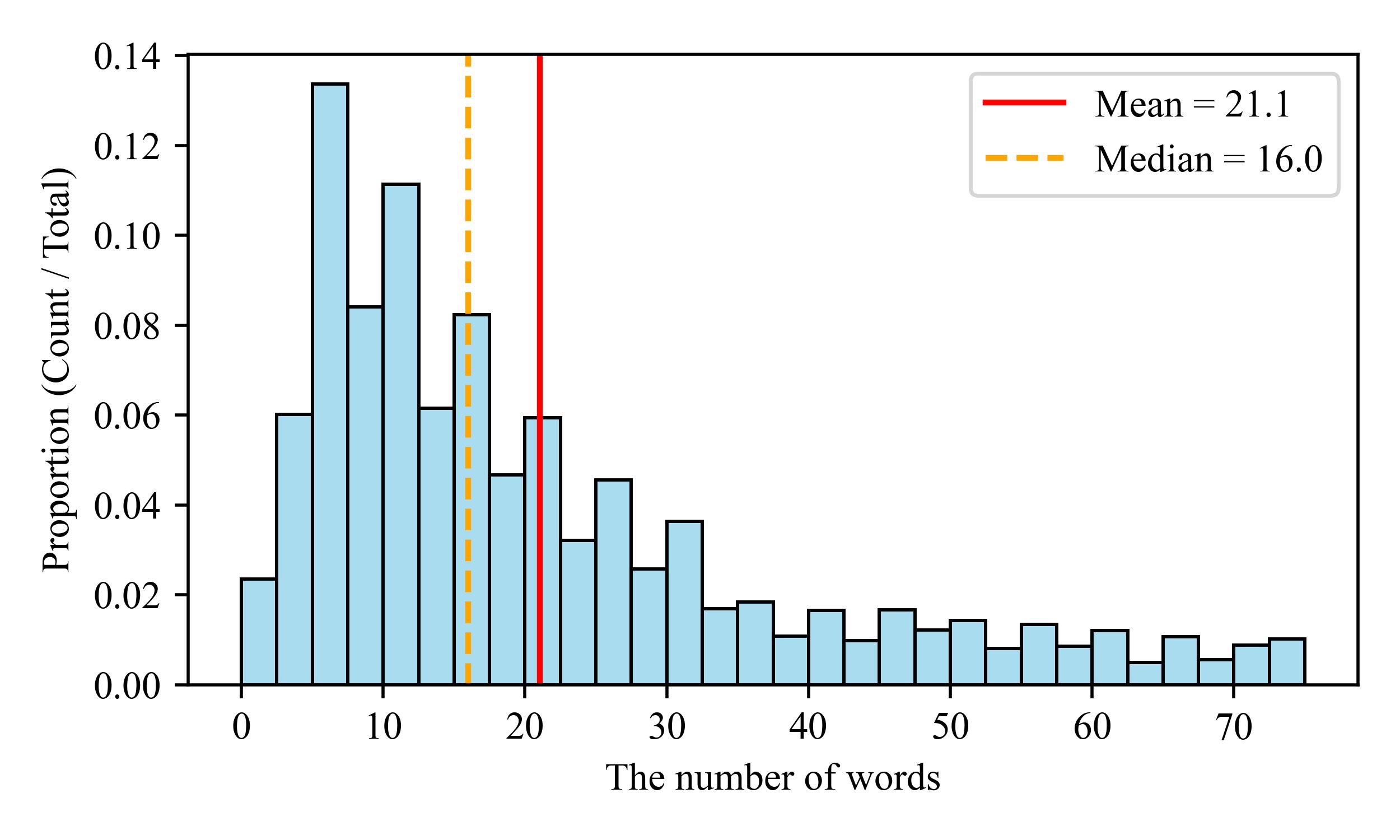}
    } 
    \caption{The word count of descriptive fields such as Title and All tags in the SMPD data.}
\label{fig:descriptive_fields}
\end{figure}

Current social media popularity prediction methods exhibit some limitations. At the VLM application level, ablation studies on the BLIP-based HVLA framework indicate that the contribution of the visual encoder alone accounts for only 65\% of the full model performance. In terms of textual representation, existing methods typically construct textual prompts using the \textit{Title} and \textit{All tags} fields, which are then aligned with the visual features. However, as shown in Figures~\ref{fig:descriptive_fields}, and supported by exploratory data analysis, the \textit{Title} field in the Social Media Prediction Dataset (SMPD) provides extremely limited information—some samples contain only a single word or even empty strings. Although the \textit{All tags} field appears to contain richer textual information, we observe that it also includes a substantial amount of vague or ambiguous content. These findings not only highlight the insufficiency of current visual feature extraction but also help explain the underlying causes of the semantic misalignment between visual and textual modalities. Methodologically, most existing approaches overly rely on image-text matching at the individual post perspective while neglecting the cross-content relational patterns inherently present in social media. This leads to an inability to capture the hierarchical propagation characteristics crucial to accurate prediction. For instance, visually similar posts (cosine similarity $>$ 0.85) from the same creator can exhibit popularity differences of up to 300\%, whereas certain categories (e.g., ``food'') maintain relatively stable popularity distributions (standard deviation $<$ 0.15) across different users. These seemingly contradictory observations underscore the presence of systematic biases in current methods’ understanding of social media content propagation dynamics.

More importantly, we observe a significant correlation between prediction error and the alignment quality of features extracted by vision-language models, particularly in terms of image-text consistency. 
This finding reveals a fundamental challenge in multimodal learning: the quality of cross-modal alignment is a key determinant of downstream task performance. 
To address above issues, we innovatively formulate a classification task based on the SMPD and fine-tune the CLIP pre-trained model accordingly. Furthermore, we propose a global-local dual-grained category-aware prompt learning framework to enhance semantic alignment at both coarse and fine levels. The main contribution of the method is as follows:

\begin{itemize}
\item \textbf{Classification-based Vision-Text Alignment}: We establish a 77-class SMPD classification task using frozen CLIP features, enhancing their discriminability through training. To overcome CLIP's visual representation limits and cross-content correlation challenges, we introduce hierarchical category prototypes and contrastive vision-text alignment.

\item \textbf{Fine-grained Semantic Alignment Strategy}: We design learnable global and local prompt templates to guide and align textual information at different granularities, thereby mining more detailed and enriched category semantic features.

\item \textbf{Cross-modal Feature Enhancement Mechanism}: We propose a cross-modal interaction module based on multi-head attention that dynamically fuses visual, textual, and prototype features, effectively improving semantic alignment and feature discriminability across modalities, enabling each sample to express more complete hierarchical semantic information.

% \item \textbf{Lightweight Popularity Prediction System}: Considering the computational cost of VLM feature extraction and tree-based ensemble regression, we propose a lightweight training strategy incorporating high-value sample selection. Representative important samples identified during the classification phase are used to train efficient tree models, achieving a robust and efficient social media popularity prediction system without significant performance loss.
\end{itemize}

\section{Related Work}
\subsection{Social Media Popularity Prediction}
Social Media Popularity Prediction (SMPP) \cite{SMP3} has become a crucial area of research due to the exponential growth of user-generated content and dynamic attributes. 
Researchers have developed various models and techniques to predict the popularity of social media posts, focusing on different aspects such as feature extraction, user behavior, and network dynamics \cite{SMP4,SMP5,SMP6}.
Some methods integrate multi-modal feature extraction for feature generalization and temporal modeling \cite{RW4,RW5,RW6}. 
Other approaches deeply exploit various types of features, including visual and language information, to improve the reliability in modeling post popularity. These methods analyze images from multiple perspectives while extracting semantic embeddings from diverse textual sources like titles, tags, concepts, and categories. Recent methods further improve prediction accuracy by utilizing contrastive learning between titles and tags while also embedding user identification features. In addition, some approaches introduce diverse feature mining techniques and stacking modules to extract informative patterns from both text and images.

\subsection{Representations with Vision-Language Models}
The effectiveness of popularity prediction relies on the advanced capabilities of multimodal representation, which accurately represents complex data across multiple types of modalities \cite{CHENG1}. Vision-language representation learning covers a wide range of tasks, including visual question answering, image captioning, image-text matching, and visual reasoning \cite{CHENG2,CHENG3}. Progress in natural language processing, image recognition, and other related technologies continues to drive this field forward. Some frameworks are designed to learn unified representations of images and text using adversarial training, enabling cross-modal translation and alignment. Others extend transformer-based language models to multimodal contexts, showing how pretraining benefits diverse vision-and-language tasks. Notably, large-scale pretraining models based on contrastive learning between image and text pairs have significantly improved performance \cite{VLM1,VLM2,ge1,ge2}. Building on these models, newer systems combine frozen pre-trained image encoders with large language models to further enhance vision-language understanding and generation. For example, Wu ~\textit{et~al.} ~\cite{RW1} introduced a novel vision-language fusion method combining multi-aspect visual analysis (semantics/quality/scenes) and {multi-granularity textual embedding (word/sentence-level), advancing social media popularity prediction. 
Chen~\textit{et~al.} proposed TTC-VLT~\cite {RW2}, solving label-title inconsistency and user feature transformation via title-tag contrastive learning combined with user identity embedding. 
Mao~\textit{et~al.} ~\cite{RW3} introduced a high-order feature mining framework with \textit{stacking blocks} to extract deep representations (beyond low-order embeddings) from multimodal data for social media popularity prediction.

The Contrastive Language-Image Pretraining (CLIP) \cite{CLIP} framework consists of an image encoder $f_I(\mathbf{x})$ and a text encoder $f_T(\mathbf{t})$, which project visual and textual inputs into a shared embedding space. The image encoder can be implemented using either ResNet \cite{Resnet} or Vision Transformer (ViT) \cite{ViT}, while the text encoder employs a Transformer-based design. 
During pretraining, CLIP learns multimodal representations through contrastive learning on 400 million image-text pairs, optimizing the alignment between matching pairs while separating non-matching ones in the embedding space. For feature extraction tasks, the framework provides direct access to the encoded representations: visual features are obtained as $\mathbf{v} = f_I(\mathbf{x})$ for input images, and textual features as $\mathbf{w} = f_T(\mathbf{t})$ for input text, with both modalities mapped to a unified semantic space that enables cross-modal similarity computation. This architecture serves as a versatile foundation for multimodal feature extraction across various downstream applications.

\section{Proposed Method}
We propose an efficient cross-modal feature extraction framework that integrates prototype representations with attention mechanisms to achieve deep feature-text alignment through classification-based training. For fine-grained text-text alignment, we design learnable dynamic prompt templates composed of global and local context vectors, constructing a semantic template library to improve intra-modal semantic consistency. For vision-text alignment, the visual prototype network maps image region features into a shared semantic space, where multi-head attention dynamically aligns visual regions with text tokens under supervision.

\subsection{Preliminaries}
Although CLIP achieves strong performance in general scenarios, it underutilizes visual information, which remains a key limitation. Its heavy reliance on image-text matching for zero-shot tasks can pose risks in certain cases. As noted in~\cite{Aircraft,UCF101}, some datasets present challenges—for example, FGVCAircraft uses class names like "737-200" that lack meaningful semantics, while UCF101 contains video frame images that poorly align with prompts such as "a photo of a {class}". The SMPD exhibits similar issues, with detailed analysis provided in Experiment section Figure \ref{fig:Detial_analyse}.

Motivated by this intuition, we construct 256-shot multimodal prototypes to represent each semantic class through both visual and textual modalities. 
For visual prototypes, we compute the mean feature vector from the CLIP image encoder $f_I$ using 256 representative images per class, denoted as $\mathbf{V}i = \frac{1}{256} \sum_{n=1}^{256}\mathbf{x}_n$ where $\mathbf{x}_n$ represents the visual feature of the $n$-th image in the $i$-th class. 
In parallel, we build textual prototypes by encoding 256 diverse class-descriptive titles through the CLIP text encoder $f_T$, formulated as $\mathbf{T}_i = \frac{1}{256} \sum_{n=1}^{256} f_T(\mathbf{t})$, where each title varies in stylistic attributes while preserving class semantics. Instead of using 11 categories, we adopt 77 ($K$) fine-grained subcategories to construct the prototypes, which significantly enhance the discriminative power of the resulting features.
To ensure that each prototype captures diverse and representative features of its corresponding class, we adopt a diversity-aware sampling strategy with a three-stage filtering process, considering temporal, semantic, and user-level variations:

\textbf{Temporal Diversity.} Social media content often changes over time due to evolving trends and seasonal events. To mitigate temporal bias, we first divide the full timeline into several intervals and sample images proportionally from each period. This ensures temporal coverage and prevents the prototypes from being dominated by short-term bursts.

\textbf{Semantic Diversity via Subtopic Filtering.} Within each temporal segment, we further filter images according to their associated third-level subtopics. By covering multiple subtopics under the same class label, we enhance the semantic completeness and richness of the prototype.

\textbf{User Diversity.} Finally, to avoid prototype bias toward individual posting styles or content preferences, we ensure that the selected images come from diverse users. This helps the prototype generalize better across different user behaviors and aesthetic tendencies.

\begin{figure*}[!h]
\centering
\includegraphics[width=0.95\textwidth]{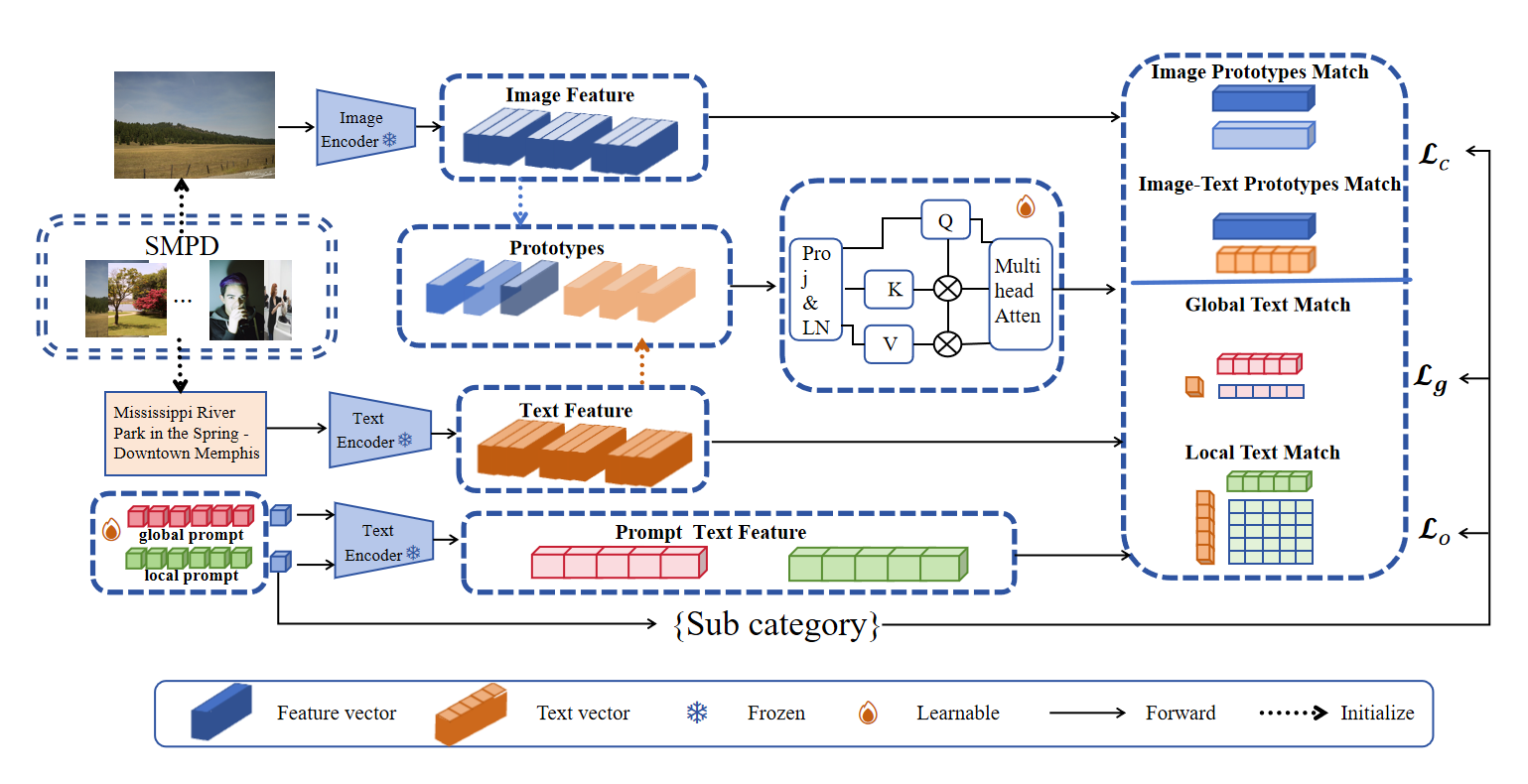} 
% \caption{During training, we use two identical text encoders from pre-trained CLIP to extract the global and local class embeddings (denoted as $\mathbf{G}$ and $\mathbf{L}$) and the overall and sequential text embeddings (denoted as $\mathbf{h}$ and $\mathbf{H}$) from the prompts and text description.}
\caption{The framework of multi-class classification training based on the SMPD.}
\label{fig:overview}
\end{figure*}

\subsection{Fine-grained Textual Alignment}
Traditional VLMs typically perform feature alignment between an entire image and its paired text (e.g., title or description) to achieve multimodal semantic fusion. However, in SMPP tasks, mainstream methods often adopt template-based prompt designs, such as ``a picture/post with title \{title\}'' or ``a picture/post belonging to class \{category\}''. 
These prompt strategies suffer from notable limitations. On the one hand, many images or posts lack informative titles, resulting in incomplete or insufficient textual inputs. On the other hand, social media images often contain complex and diverse semantics, and simple prompt templates built from short titles fail to capture fine-grained details and high-level concepts within the image.

To address these challenges, we propose a \textit{fine-grained textual alignment strategy} built upon a classification framework. Instead of relying on global text-image matching, our approach structurally models key textual components, such as attribute words, semantic fragments, and contextual information, and performs fine-grained alignment with corresponding regions or semantic subspaces in the image. This strategy enhances the model's robustness in scenarios with incomplete text or semantically complex visual content.

\begin{table*}[!h]
\centering
\caption{Experimental results with various settings. \textbf{Bold} indicates the best score. $\uparrow$ indicates higher is better, $\downarrow$ indicates lower is better.}
\label{tab:results}
\resizebox{0.95\textwidth}{!}{ 
\begin{tabular}{ccccccccccccc}
\toprule
\multirow{2}{*}{Statistic feature} & \multirow{2}{*}{User Behavior feature} & \multirow{2}{*}{Resnet+Bert} & \multirow{2}{*}{ViT+Bert} & \multirow{2}{*}{CLIP} & \multirow{2}{*}{BLIP} & \multirow{2}{*}{CLIP$_{\text{FT}}$} & \multicolumn{2}{c}{LightGBM} & \multicolumn{2}{c}{CatBoost} & \multicolumn{2}{c}{Fusion} \\
& & & & & & & SRC $\uparrow$ & MAE $\downarrow$ & SRC $\uparrow$ & MAE $\downarrow$ & SRC $\uparrow$ & MAE $\downarrow$ \\
\midrule
\checkmark & & & & & & & 0.7023 & 1.2783 & 0.6833 & 1.2971 & 0.7042 & 1.2615 \\
\checkmark & \checkmark & & & & & & 0.7173 & 1.2562 & 0.7077 & 1.2464 & 0.7119 & 1.2301 \\

\checkmark & & \checkmark & & & & & 0.7219 & 1.2318 & 0.7058 & 1.2520 & 0.7223 & 1.2214 \\
\checkmark & & &\checkmark& & & & 0.7223 & 1.2322 & 0.7135 & 1.2464 & 0.7285 & 1.2148 \\
\checkmark & & & & \checkmark& & & 0.7250 & 1.2249 & 0.7161 & 1.2353 & 0.7296 & 1.2087 \\
\checkmark & & & & & \checkmark& & 0.7241 & 1.2283 & 0.7157 & 1.2354 & 0.7292 & 1.2093 \\
\checkmark &  & & & & & \checkmark & 0.7278 & 1.2052 & 0.7290 & 1.2062 & 0.7330 & 1.1979 \\

\checkmark & \checkmark & \checkmark& & & &  & 0.7245 & 1.2208 & 0.7110 & 1.2414 & 0.7276 & 1.2278 \\
\checkmark & \checkmark & &\checkmark & & & & 0.7258 & 1.2215 & 0.7170 & 1.2407 & 0.7297 & 1.2210 \\
\checkmark & \checkmark & & & \checkmark& & & 0.7225 & 1.2287 & 0.7112 & 1.2448 & 0.7258 & 1.2154 \\
\checkmark & \checkmark & & & & \checkmark& & 0.7237 & 1.2264 & 0.7127 & 1.2135 & 0.7269 & 1.2036 \\
% \checkmark & \checkmark & & & & &\checkmark & 0.7226 & 1.2354 & 0.7184 & 1.2343 & 0.7294 & 1.2123 \\
\checkmark & \checkmark & & & & & \checkmark & 0.7295 & 1.2007 & 0.7303& 1.1941 & \textbf{0.7395} & \textbf{1.1810}\\
\bottomrule
\end{tabular}
}
\end{table*}

Following \cite{TaI-DPT}, we leverage two parallel copies of the pre-trained CLIP text encoder ($f_T$) to independently encode prompt templates and textual content. To capture class-level semantics at multiple granularity levels, we introduce trainable global and local prompts, which enable the extraction of both global and regional class embeddings. For the $i$-th class:
\begin{equation}
\mathbf{w}_i^G = [\theta^G_{1:s}, e_i], \quad \mathbf{w}_i^L = [\theta^L_{1:s}, e_i]
\end{equation}
where $\theta^G,\theta^L$ represent learnable embeddings concatenated with class label embeddings. 
Then $\mathbf{w}_i^G$ and $\mathbf{w}_i^L$ prompts are passed through the $f_T$ to generate global and local class embeddings:
\begin{equation}
\{\mathbf{G}_i,\mathbf{L}_i\} = f_T(\mathbf{w}_i^{G},\mathbf{w}_i^{L}) 
\end{equation}
To better preserve fine-grained regional characteristics in textual inputs, we retain the complete sequence of token embeddings (as opposed to using only the ending \texttt{<EOS>} token) and define:
\begin{equation}
\{\mathbf{h}, \mathbf{H}\} = f_T(\textbf{r})
\end{equation}
where $\textbf{r}$ denotes the input textual description (i.e., the ``Title'' or
``All tags''). 
Let $\mathbf{h} \in \mathbb{R}^d$ represent the global textual embedding, and $\mathbf{H} \in \mathbb{R}^{l \times d}$ denote the sequence of token-level embeddings corresponding to the description of length $l$. Then, the global similarity is computed by
\begin{equation}
p_i = \text{cos}\langle \mathbf{h}, \mathbf{G}_i \rangle 
\end{equation}
Further, the local branch information is spatially aggregated as follows:
\begin{equation}
p'_i = \frac{\sum_{j=1}^l \exp(P_{ij} / \tau_s)}{\sum_{j=1}^l \exp(P_{ij} / \tau_s)} \cdot P_{ij}, \quad \text{where} \quad P_{ij} =  \text{cos}\langle \mathbf{H}_j, \mathbf{L}_i \rangle 
\end{equation}
where \( \tau_s \) controls the focus on specific locations. 
The global similarity \( p_i \) and local similarity \( p'_i \) are optimized by the loss terms \( \mathcal{L}_{\text{G}} \) and \( \mathcal{L}_{\text{O}} \), respectively.

\subsection{Cross-Modal Projection}
To enable more effective feature interaction, we introduce the Cross-Modal Projection module. Firstly, we establish aligned feature spaces through dedicated projection layers: the visual projection layer $P_I$ and textual projection layer $P_T$, which map both image and text features into a unified $d$-dimensional embedding space, enabling direct cross-modal feature comparison via cosine similarity. 
To capture cross-modal interactions, we apply a multi-head self-attention mechanism to the concatenated embeddings of image features, visual prototypes, and textual prototypes. Formally:
\begin{equation}
[\tilde{\mathbf{x}}, \tilde{\mathbf{V}}, \tilde{\mathbf{T}}] = \text{SelfAttn}(\mathbf{ \text{Concat}(P_I(\mathbf{x}),P_I(\mathbf{V}), P_I(\mathbf{T}))})\
\end{equation}
This operation captures contextual relationships across modalities while maintaining permutation invariance, enabling us to recover the updated representations via output slicing. 
After acquiring the enhanced feature representations, the prediction can be performed using both visual and textual prototypes through the following probabilistic formulations:
\begin{equation}
\begin{aligned}
p_V(y_i=1) &= \frac{\exp\left(\text{cos}\langle \tilde{\mathbf{x}}, \tilde{\mathbf{V}}_i\rangle / \tau_v\right)}{\sum_{j=1}^K \exp\left(\text{cos}\langle \tilde{\mathbf{x}}, \tilde{\mathbf{V}}_j\rangle / \tau_v\right)} \\
p_T(y_i=1) &= \frac{\exp\left(\text{cos}\langle\tilde{\mathbf{x}}, \tilde{\mathbf{T}}_i\rangle  / \tau_t\right)}{\sum_{j=1}^K \exp\left(\text{cos}\langle \tilde{\mathbf{x}}, \tilde{\mathbf{T}}_j' \rangle / \tau_t\right)}
\end{aligned}
\label{eq:visual_text}
\end{equation}
where $\tau_v$ and $\tau_t$ represent the temperature parameters for the visual and textual modalities respectively.
The final prediction is obtained by combining these two modality-specific probabilities, thereby achieving effective multimodal information integration.

\subsection{Loss Design}
% We optimize groups of logits computed from multi-modal features using the softmax function. 
% Specifically, $\mathbf{z}_P$ denotes the prototype matching features, $\mathbf{z}_V$ represents the visual prototype features, $\mathbf{z}_T$ is the textual prototype features, $\mathbf{z}_G$ corresponds to the global textual matching features, and $\mathbf{z}_O$ refers to the local textual matching features. 
The overall loss function used during training is defined as:
\begin{equation}
\mathcal{L} = \mathcal{L}_{g} + \mathcal{L}_{o}+\mathcal{L}_{c}
\end{equation}
The term $\mathcal{L}_c$ denotes the alignment loss between visual features and prototypes, which is computed by applying cross-entropy to the averaged logits of $p_V$ and $p_Z$ as defined in Eq.~\eqref{eq:visual_text}. The terms $\mathcal{L}_g$ and $\mathcal{L}_o$ correspond to the global and local textual alignment losses, respectively, and are similarly optimized by applying cross-entropy to their corresponding logits.

\section{Experiment}

\subsection{Experiment Setup}

\subsubsection{Dataset}
The \textbf{SMPD-Image} dataset is a large-scale multimodal collection of social media images, comprising 486,000 image posts from 70,000 distinct users. This dataset provides not only the original visual content but also rich metadata including: anonymized user sharing records, user profile information, associated web images, textual descriptions, timestamps, geolocation data, and category labels. All data were collected from Flickr, one of the world's leading photo-sharing platforms, ensuring authentic diversity and temporal characteristics of real-world social media content.

\subsubsection{Evaluation Metrics}
We use two metrics to evaluate model performance: Spearman's Rho (SRC) and Mean Absolute Error (MAE). SRC assesses the rank correlation between predicted and true popularity, defined as
\begin{equation}
 \text{SRC} = 1 - \frac{6\sum_{i=1}^{n}d_i^2}{n(n^2 - 1)}   
\end{equation}
where $d_i$ is the rank difference and $n$ the number of samples. 

MAE measures the average absolute prediction error:
\begin{equation}
\text{MAE} = \frac{1}{n}\sum_{i=1}^{n}|y_i - \hat{y}_i|,
\end{equation}
where $y_i$ and $\hat{y}_i$ are ground-truth and predicted values respectively.

\subsubsection{Implementation Details} 
All experiments were conducted using the PyTorch framework with distributed computing acceleration on NVIDIA A100 Tensor Core GPUs. 
% For CLIP-based visual feature extraction, input images were preprocessed to a standardized resolution of $224\times224\times3$ pixels. 
We employ the AdamW optimizer with a base learning rate of $1\times10^{-4}$, weight decay of $1\times10^{-5}$, trained for 4 epochs using a batch size of 128. After completing training, we save the model parameters and perform forward inference to extract features. 
We employ a classification-optimized CLIP model to extract multimodal features, including image features, prototype features, hierarchical text prompt features (both global and local), and user behavior features $\mathbf{s}$ from~\cite{HVLA}.
\begin{equation}
\mathbf{F} = [\mathbf{f}_I(\mathbf{x}),\mathbf{f}_I(\mathbf{t}),\mathbf{w}_i^G,\mathbf{w}_i^L, \tilde{\mathbf{x}}, \tilde{\mathbf{V}}, \tilde{\mathbf{T}},\mathbf{s}]
\end{equation}

\subsection{Results and Analysis}
To evaluate our method, we adopt LightGBM~\cite{LightGBM} and CatBoost~\cite{CatBoost} as regression models for post popularity prediction. Our proposed fine-tuned CLIP strategy (CLIP$_{\text{FT}}$) achieves the best performance, significantly outperforming features extracted from standard pre-trained CLIP and BLIP2. As shown in Table~\ref{tab:results}, CLIP$_{\text{FT}}$ achieves an SRC of \textbf{0.7295} with LightGBM and \textbf{0.7303} with CatBoost. In addition, user behavior features—such as tag sequences and tag frequency—serve as latent indicators of post popularity, supporting the hypothesis of a strong correlation between user characteristics and content virality. The integration of these hand-crafted user features with aligned multimodal representations further enhances model performance, with LightGBM consistently outperforming CatBoost on the SMPD. Moreover, ensemble learning substantially improves robustness beyond individual model capabilities, addressing the inherent limitations of single model approaches and yielding notable performance gains.

As illustrated in Figure~\ref{fig:Distribution}, we examine the SRC prediction error for tail samples with popularity scores exceeding 12. A simple oversampling strategy \cite{oversampling} is applied, which effectively reduces local prediction errors on the validation set. However, it fails to improve the overall SRC performance. We hypothesize that this outcome results from a dual effect: model overfitting to tail samples and a systematic prediction bias caused by a rightward shift in the output distribution. We observe that categories such as \textit{animal}, \textit{food}, and \textit{electronic} exhibit high image-text alignment, often yielding higher average SRC and lower MAE in prediction. In contrast, categories like \textit{weather} and \textit{season} demonstrate weaker alignment, which is intuitive: compared to the more concrete concepts in \textit{animal} and \textit{food}, images in \textit{weather} and \textit{season} are often more abstract and visually complex, making it harder for the model to establish accurate semantic correspondence. As a result, these categories tend to have lower SRC and higher MAE. This phenomenon indicates that the degree of semantic alignment between image and text significantly impacts downstream modeling and prediction performance. Higher alignment enables the model to learn more discriminative multimodal representations, leading to improved predictive accuracy.

\begin{figure}[!h]
\centering
\includegraphics[width=0.45\textwidth]{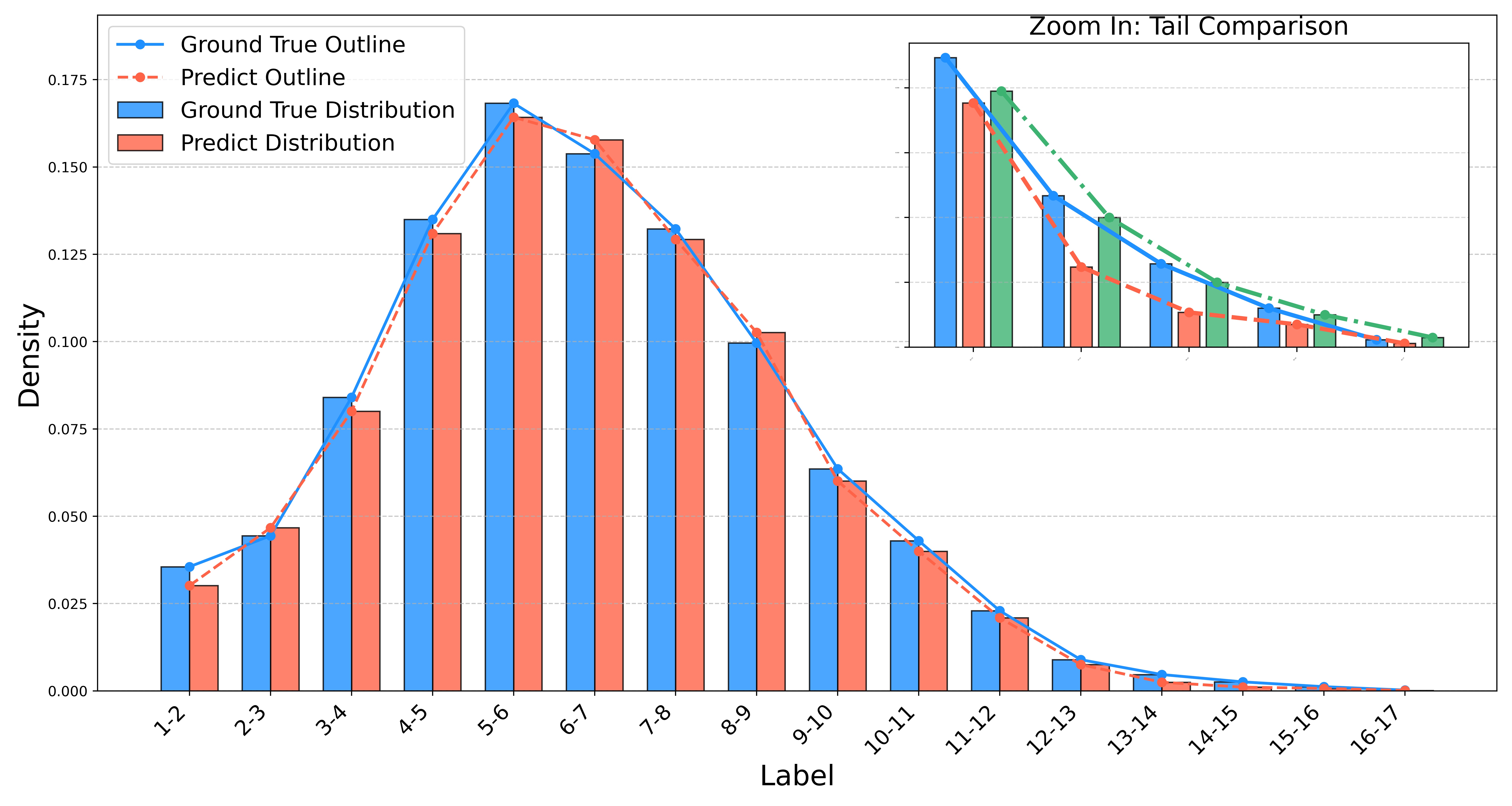} 
\caption{Distribution visualization of actual popularity vs. predicted popularity.}
\label{fig:Distribution}
\end{figure}

\begin{figure}[!h]
    \centering
    \subfigure[Distribution analysis of average cosine similarity vs. average predicted metric per class in SMPD on pre-trained CLIP.]{\includegraphics[width=0.24\textwidth]{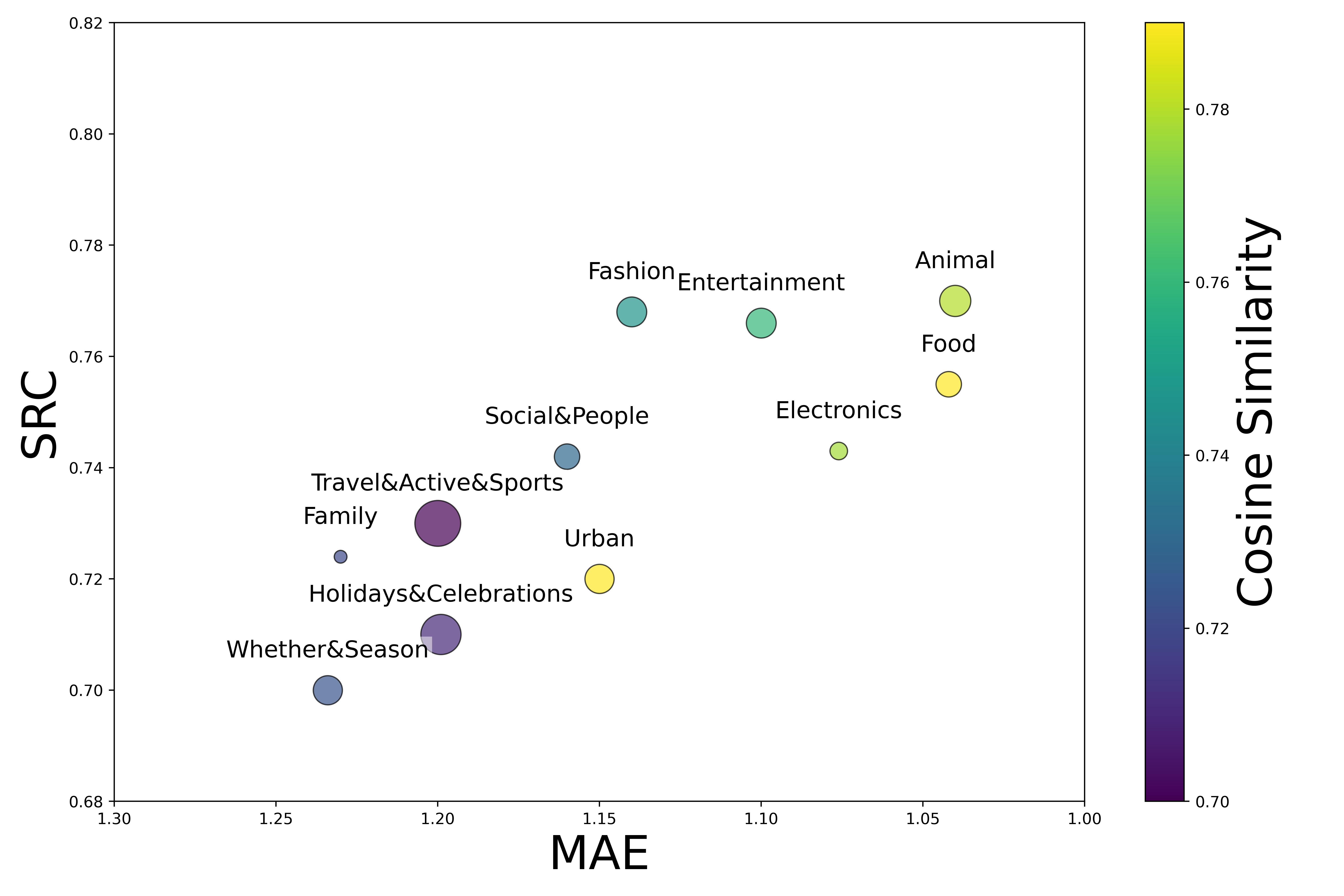}
    }
    \hspace{0.2cm} % 增加1厘米水平间距
    \subfigure[Comparison of SRC performance across varying training set ratios.]{\includegraphics[width=0.2\textwidth]{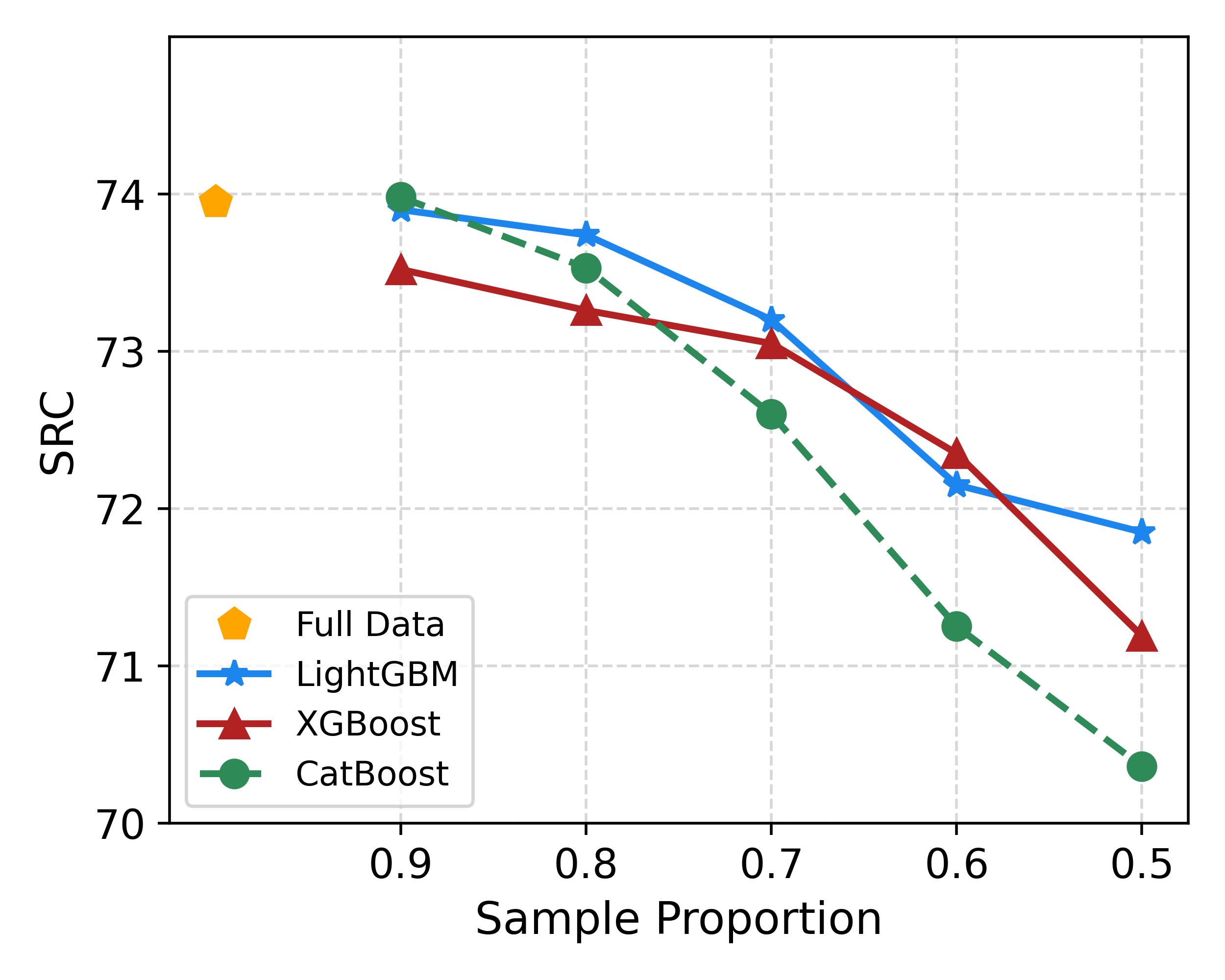}
    } 
    \caption{The word count of descriptive fields such as Title and All tags in the SMPD data.}
\label{fig:Detial_analyse}
\end{figure}

% Moreover, the volume of training data has long posed a challenge for ensemble tree models, primarily due to their high computational cost. To address this, we incorporate a loss-based sample selection strategy. Experimental results show that using only 77\% of the training data based on LightGBM does not lead to a significant drop in SRC, suggesting that a carefully filtered subset can maintain performance while reducing training overhead.

Traditional ensemble tree models consistently face computational efficiency challenges due to excessive training data requirements. To address this, we propose an innovative loss-based sample selection mechanism that filters training samples by ranking their supervised classification losses. Experimental results under the LightGBM framework demonstrate that using only 77\% of selected training samples maintains stable SRC performance (with less than 1\% degradation) while significantly reducing computational overhead.

\subsection{Discussion}
\textit{Q1: Why combine Vision-Language Models with tree-based ensemble learning methods?}

A widely adopted approach in multi-modal learning pipelines is to use pretrained Vision-Language Models, such as CLIP or BLIP, as frozen feature extractors to obtain semantically rich and transferable embeddings from image-text pairs. These embeddings can then be treated as structured inputs for traditional classifiers. In our framework, we follow this paradigm by combining VLM-based feature extraction with LightGBM, a gradient-boosted decision tree model that is well-suited for tabular data and provides strong performance with good interpretability. This design is computationally efficient, avoids costly fine-tuning, and enables flexible and scalable multi-modal classification.

\textit{Q2: How do current social media popularity prediction methods fail to adequately incorporate temporal features in their analyses?}

Current SMPP models tend to systematically overlook temporal dynamics, which constitutes a critical limitation given the inherently time-sensitive nature of user interactions on social platforms. While existing approaches excel in analyzing multimodal content, their treatment of temporal features remains static—typically relying on hand-crafted statistics from the user perspective within the time dimension—which fails to capture the underlying rhythms of content propagation, such as platform-specific decay patterns and circadian fluctuations in user engagement. In our experiments, we attempted to introduce temporal modeling mechanisms, but the results were suboptimal. Beyond the limitations of the data itself (e.g., timestamp noise), we also surveyed related tasks, such as the Kaggle Jane Street financial prediction competition and the China Collegiate Big Data Challenge on meteorological forecasting. These investigations revealed that the prevailing paradigm still centers around feature engineering combined with regression-based models, as outlined in \textit{Q1}.

% \textit{Q: \textcolor{red}{Modeling strategies from different perspectives.}}
% Existing approaches typically predictions based on the minimal unit of granularity, i.e., individual posts or photos, and adopt a modeling strategy that combines feature extraction with ensemble tree-based models. However, this paradigm exhibits notable limitations, as transforming raw data into structured tabular formats inevitably leads to the loss of two critical types of information. First, temporal dynamics are substantially diminished. Rich temporal patterns inherent in the raw data—such as the distribution of user posting intervals or the evolution curves of content popularity—are often oversimplified during feature engineering. This simplification typically manifests as compressing continuous time series into scalar statistics like mean or variance, while neglecting platform-specific temporal behaviors and periodicities. Second, contextual information related to user behavior is difficult to preserve effectively. In particular, features derived from social network structures and long-term interest drift—evident from the semantic evolution of a user's historical posting topics—are often lost. These deeper aspects of user profiling cannot be fully captured through isolated, static snapshots of individual samples. 

\section{Conclusion}
% This paper constructs a multi-class classification task based on the Social Media Prediction Dataset, proposes hierarchical category prototypes and a contrastive learning strategy to enhance semantic alignment, and designs a global-local dual-grained prompt learning framework along with a multi-head attention cross-modal interaction module to achieve dynamic fusion of visual, textual, and prototype features. Extensive experiments demonstrate the superiority of the proposed method. However, current techniques still struggle to capture the complex temporal dynamics inherent in social media content propagation. Future research in social media popularity prediction should, while maintaining strong vision-language representation capabilities, deeply integrate dynamic temporal features and user behavior patterns, promoting the collaborative advancement of multimodal semantic alignment and temporal dynamic modeling to improve prediction accuracy and generalization.
This paper proposes a novel framework for social media popularity prediction, introducing hierarchical category prototypes with contrastive learning to enhance semantic alignment, and developing a dual-grained prompt learning approach with multi-head attention for cross-modal feature fusion. Extensive experiments demonstrate our method's superior performance, though capturing complex temporal dynamics in content propagation remains challenging. Future research should integrate temporal dynamics and user behavior patterns while maintaining robust vision-language representations. This dual focus on multimodal alignment and temporal modeling will advance prediction accuracy and generalization. 
% Future research in social media popularity prediction should, while maintaining strong vision-language representation capabilities, deeply integrate dynamic temporal features and user behavior patterns.

\section{Acknowledgments}
This work is supported by the National Natural Science Foundation of China under Grants Nos. 62441225, 61972192, 62172208, 61906085.
This work is partially supported by Collaborative Innovation Center of Novel Software Technology and Industrialization.
This work is supported by the Fundamental Research Funds for the Central Universities under Grant No. 14380001.

\bibliographystyle{ACM-Reference-Format}
% \bibliography{ref}

\end{document}